\documentclass[
 aps, pra,
 amsmath,amssymb,
 11pt,
 final,
tightenlines,
 twoside,
 twocolumn,
 nofloats,
nofootinbib,
 superscriptaddress,
showkeys,
showkeywords,
 ]
{revtex4-2}

\usepackage[T2A]{fontenc}
\usepackage[utf8x]{inputenc}
\usepackage[russian,english]{babel}
\usepackage{graphicx}
\usepackage{dcolumn}
\usepackage{bm}
\usepackage{xcolor}

\newcommand{\CII}{[C\,{II}]}
\newcommand{\hii}{H\,{II}}
\newcommand{\HI}{H\,{I}}

\newcommand{\brg}{Br$\gamma$}
\newcommand{\feii}{[Fe\,{II}]}
\newcommand{\micron}{$\mu$m}
\input{maik.rty}

\setcitestyle{authoryear,round}
\def\squareforqed{\hbox{\rlap{$\sqcap$}$\sqcup$}}

\def\sq{\ifmmode\squareforqed\else{\unskip\nobreak\hfil
\penalty50\hskip1em\null\nobreak\hfil\squareforqed
\parfillskip=0pt\finalhyphendemerits=0\endgraf}\fi}

\def\arcsec{\hbox{$^{\prime\prime}$}}

\def\utw{\smash{\rlap{\lower5pt\hbox{$\sim$}}}}

\def\udtw{\smash{\rlap{\lower6pt\hbox{$\approx$}}}}

\def\diameter{{\ifmmode\mathchoice
{\ooalign{\hfil\hbox{$\displaystyle/$}\hfil\crcr
{\hbox{$\displaystyle\mathchar"20D$}}}}
{\ooalign{\hfil\hbox{$\textstyle/$}\hfil\crcr
{\hbox{$\textstyle\mathchar"20D$}}}}
{\ooalign{\hfil\hbox{$\scriptstyle/$}\hfil\crcr
{\hbox{$\scriptstyle\mathchar"20D$}}}}
{\ooalign{\hfil\hbox{$\scriptscriptstyle/$}\hfil\crcr
{\hbox{$\scriptscriptstyle\mathchar"20D$}}}}
\else{\ooalign{\hfil/\hfil\crcr\mathhexbox20D}}%
\fi}}

\begin{document}

\selectlanguage{english}

\keywords{\it galaxies: abundances, galaxies: individual: ISM}

\title{Near infrared view on the photodissociation regions S\,255, S\,257, NGC\,7538 and S\,140}

\author{\firstname{M.~S.}~\surname{Kirsanova} }
\affiliation{Institute of Astronomy, Russian Academy of Sciences, Pyatnitskaya str. 48, Moscow 119017, Russia}

\author{\firstname{A.~M.}~\surname{Tatarnikov} }
\affiliation{M.~V. Lomonosov Moscow State University, Sternberg Astronomical Institute, 119991, Universitetskij pr.~13, Moscow, Russia} 

\author{\firstname{P.~A.}~\surname{Boley} }
\affiliation{Institute of Astronomy, Russian Academy of Sciences, Pyatnitskaya str. 48, Moscow 119017, Russia}

\author{\firstname{D.~S.}~\surname{Wiebe} }
\affiliation{Institute of Astronomy, Russian Academy of Sciences, Pyatnitskaya str. 48, Moscow 119017, Russia}

\author{\firstname{N.~A.}~\surname{Maslennikova} }
\affiliation{M.~V. Lomonosov Moscow State University, Sternberg Astronomical Institute, 119991, Universitetskij pr.~13, Moscow, Russia} 

\author{\firstname{A.~A.}~\surname{Tatarnikov} }
\affiliation{M.~V. Lomonosov Moscow State University, Sternberg Astronomical Institute, 119991, Universitetskij pr.~13, Moscow, Russia} 

\begin{abstract}

We performed photometric observations of the S\,255, S\,257, S\,140, NGC\,7358 and the Orion~Bar photo-dissociation regions (PDRs) at 2~$\mu$m using narrow-band filters centered on the \brg, H$_2$ and \feii{} lines, as well as the narrow-band Kcont and the broad-band $H$ filters for continuum subtraction. The observations were done with the 2.5-m telescope of the SAI Caucasian Mountain Observatory and the near-infrared camera and spectrograph ASTRONIRCAM. We find several high-density arc-like structures in the \brg{} and \feii{} images of the ionized gas in NGC\,7538 and extended shells and arcs visible through the H$_2$ emission. The H ionization front and H$_2$ dissociation front are merged in NGC\,7538. In S\,255 and S\,257 we detected only \brg{} emission from the \hii{} regions and bright H$_2$ emission from the PDRs. The projected distance between the H ionization and H$_2$ dissociation fronts are $\approx 0.3-0.4$~pc, which cannot be explained using models of a uniform medium. Most probably, the ionized and neutral gas in these PDRs is clumpy. The \HI-to-H$_2$ transitions in the NGC\,7538, S\,255, S\,257 and S\,140 PDRs are gradual with no sharp borders. This conclusion also agrees with the suggestion of a clumpy medium.

\end{abstract}

\maketitle

\section{Introduction}\label{sec:intro}

Young massive stars signpost themselves in the infrared (IR) sky via ring-like shells \citep[e.~g.][and many others]{Churchwell_2006,2008ApJ...681.1341W, Deharveng_2010,2018ARep...62..764T}. These shells represent photo-dissociation regions (PDRs) created by far-UV photons with $h\nu < 13.6$~eV, which dissociate H$_2$, CO, O$_2$ and other molecules, as well as heat up interstellar dust \citep[e.~g.][]{Tielens_1985, Sternberg_1995}. The IR shells resemble projections of spheres in the plane of the sky, but many of them are in fact two-dimensional structures reminiscent of tori \citep[e.~g.][]{Beaumont_2010}. In some cases different tracers, such as dust emission, molecular or atomic lines, show different shell geometry due to non-uniform gas distribution and excitation conditions \citep[e.~g. series of works about \hii{} region RCW~120 by][]{Deharveng_09, Pavlyuchenkov_2013, Anderson_2015, Marsh_2019, Kirsanova_2019_rcw120, 2021SciA....7.9511L, 2022A&A...659A..36K}. 

In early studies \citep[e.~g.][]{Tielens_1985b} photo-dissociation regions (PDRs), separating an \hii{} region and a molecular cloud, were considered as structures in equilibrium. However, subsequent simulations showed that dynamical effects can be responsible for such phenomena as merging of an H ionization front (IF) and an H$_2$ dissociation front (DF), as well as merging of H$_2$ and CO dissociation fronts \citep[e.~g.][]{Hosokawa_2005}. The latter was successfully demonstrated for the Orion Bar PDR by \citet{Goicoechea_2016} and \citet{Kirsanova_2019}. Non-stationary PDRs, often resembling shells, produce a continuous network of bubbles and arcs in the far-IR, while not much is known about their appearance in the near-IR, with the exception of several bright objects. For example, \citet{Walmsley_2000} studied the geometry of the Orion Bar PDR using bright emission lines of H$_2$ and various atomic species and ions in the near-IR. They achieved a spatial resolution of 0.6\arcsec, which corresponds to 0.0012~pc (or 240~AU) at a distance of 414~pc~\citep{2007A&A...474..515M}, and found a separation between the fronts of H ionization and H$_2$ dissociation of 17--20\arcsec, depending on the particular position in the PDR \citep[see also][]{1996A&A...313..633V, 2000AJ....120..382O}. This stricture was also recently confirmed by \citet{Habart_2022}. 

Inspired by the recent commissioning of the SAI Caucasian Mountain Observatory (CMO), we performed near-IR photometric observations of the S\,255, S\,257, NGC\,7538, S\,140 and the Orion~Bar PDRs to study their geometry and physical conditions.  We selected filters corresponding to the \brg, \feii{} lines from the ionized gas, and the H$_2$ lines, excited by far-UV photons in PDRs \citep{Marconi_1998, Walmsley_2000}. While the \brg{} and H$_2$ lines allowed locating the IF and DF themselves, the \feii{} line was used to find the densest parts of the ionization front, as these lines are collisionally excited. So far, only the compact areas of these PDRs, associated with bright peaks of IR or molecular emission from embedded massive protostars, have been extensively studied in the near-IR, and not much attention has been given to the structure and physical conditions of the selected PDRs. All of the selected PDRs are included in the extensive study of three-dimensional structure started in the work by \citet{Kirsanova_2020_3d}.

The main physical properties of the \hii{} regions with associated PDRs are summarized in Table~\ref{tab:known_phys_properties}. Whenever possible, we use parallax-based Gaia~DR3 distances \citep{Gaia_DR3} to the ionizing stars as the distances to the PDRs (LS\,19 in S\,255 and HD\,211880 in S\,140). For S\,257, however, the quality of the photometric solution from Gaia is still insufficient. Therefore, we use an estimate of 2.5~kpc for the distance, based on a spectrophotometric measurement of $2.46\pm0.16$~kpc obtained by \citet{Russeil_2007} (S\,257 is designated as 192.50--0.10 in this work), and several other measurements summarized by \citet{Chavarria_2008}. The spectrophotometric distance to NGC\,7538 is $2.4\pm0.14$~kpc \citep{Russeil_2007} (designated as 111.50--0.80 in this work). This value agrees with the parallax-based distance $2.65^{+0.12}_{-0.11}$~kpc to a methanol maser in the direction to the south-east of the ionizing star \citep{2009ApJ...693..406M}, and also with the kinematic distance esimate of $2.44\pm0.77$~kpc from \citet{2015AJ....150..147F} (in this work the object as designated as S\,158). Based on those measurements, we adopt 2.5~kpc as the distance to NGC\,7538 in the present study.

\begin{table*}
    \caption{Properties of the \hii{} regions towards the target PDRs. References for the parameters: [1] \citet{Ojha_2004}, [2] see Sec.~\ref{sec:intro}, [3] \citet{Ojha_2011}, [6] \citet{Kirsanova_2023_3d}, [7] \citet{Balser_2018}, [8] \citet{Gaia_DR3}, [9] this is the shortest distance between the ionizing star and the border of the PDR.}
    \centering
    \begin{tabular}{c|c|c|c|c}
         \hline
                   & S\,255 & S\,257 & NGC\,7538 & S\,140\\
         \hline
     ionizing star & O9.5~V$^3$, LS\,19 & B0.5~V [3], HD\,253327 &   O5-O6, IRS~6 [1] & B0V, HD~211880\\
     Distance      & 2060$^{2181}_{1951}$~pc [8]  &  2.5~kpc [2] &  2.5~kpc [2] & 921$\pm$16~pc [8]\\
     Size  &  210\arcsec{} [6]&  190\arcsec{} [6]& 150\arcsec & 376\arcsec{} [9]\\
     $n_e$ & $\approx 100$ [6] & $\approx 100$ [6] & 100--1000 [7] & diffuse\\
         \hline
    \end{tabular}
    \label{tab:known_phys_properties}
\end{table*}

\section{Observation and data reduction}\label{sec:observations}


Photometric observations of the S\,255, S\,257, S\,140, NGC~7358 and Orion Bar PDRs were obtained over two observing sessions in November 2020 and January 2021 using the 2.5-m telescope of the SAI Caucasian Mountain Observatory and the near-infrared camera and spectrograph \mbox{ASTRONIRCAM} \citep{Nadjip_2017}, which provides a $4.6'\times4.6'$ field of view in a single exposure. Observations were carried out at nights with seeing of 0.7--1\arcsec{} (at visible wavelengths) and stable atmospheric transmission. The astroclimate parameters were measured using a MASS-DIMM instrument \citep{tokovinin_2007}. Observations were performed using narrow-band filters for \brg{} (2.1673~$\mu$m, FWHM 21.2~nm), H$_2$ (2.132~$\mu$m, FWHM 46~nm) and \feii{} (1.6442~$\mu$m, FWHM 26.2~nm), as well as the narrow-band Kcont filter (2.2729~$\mu$m, FWHM 39.4~nm) and broad-band $H$ filter (1.4901--1.7829~$\mu$m) for continuum subtraction.  For each target field and filter, we made a series of short exposures (3--30~s, depending on the brightness of the stars in the field) and employed dithering between individual exposures to minimize the effects of bad pixels, with typical shifts of 3\arcsec{} between subsequent frames. We summarize the observations in Table~\ref{tab:obslog}, where we also list the resulting spatial resolution for each field and filter, as determined from measurements of the point spread function (PSF).

\begin{table*}
    \caption{Observing log}
    \centering
    \begin{tabular}{l|c|c|c|c|c|c|c}
    \hline
    Name & RA (J2000) & Dec (J2000) & Filter & Observation date & Total exp. time & Frame exp. time & Spatial resolution \\
         & (hms)      & (dms)       &        &                  & (s)             & (s)             & (\arcsec) \\
    \hline
     NGC\,7538 & 23:13:38 & +61:29:49 &          H$_2$ & 2020-11-27 &    788 & 29.170 & 1.09 \\
             &          &           &          Kcont & 2020-11-27 &    875 & 29.170 & 1.11 \\
             &          &           &         \brg{} & 2021-01-02 &    492 &  9.120 & 1.19 \\
             &          &           &          H$_2$ & 2021-01-02 &   1575 & 29.170 & 1.35 \\
             &          &           &          Kcont & 2021-01-02 &   1575 & 29.170 & 1.22 \\
             &          &           & \feii{} & 2021-10-26 &   1113 &  9.120 & 1.09 \\
             &          &           &              H & 2021-10-26 &    153 &  3.646 & 1.19 \\
   Orion Bar & 05:35:23 & --05:25:01 &         \brg{} & 2020-12-07 &    875 & 14.590 & 1.13 \\
             &          &           &          Kcont & 2020-12-07 &    452 & 14.590 & 1.06 \\
             &          &           & \feii{} & 2020-12-27 &   1488 & 29.170 & 1.10 \\
             &          &           &              H & 2020-12-27 &     88 &  3.646 & 1.05 \\
             &          &           &          H$_2$ & 2021-02-21 &    939 &  9.120 & 1.30 \\
             &          &           &          Kcont & 2021-02-21 &   1459 & 14.590 & 1.17 \\
        S\,140 & 22:19:21 & +63:18:56 &         \brg{} & 2020-11-27 &    447 &  9.120 & 0.96 \\
             &          &           &          H$_2$ & 2020-11-27 &   1809 & 29.170 & 1.00 \\
             &          &           &          Kcont & 2020-11-27 &   1546 & 29.170 & 0.94 \\
             &          &           & \feii{} & 2020-11-29 &   1954 & 29.170 & 0.96 \\
             &          &           &              H & 2020-11-29 &    131 &  3.646 & 0.90 \\
             &          &           &         \brg{} & 2020-12-01 &    447 &  9.120 & 1.01 \\
             &          &           &          H$_2$ & 2020-12-01 &   1021 & 29.170 & 0.89 \\
             &          &           &          Kcont & 2020-12-01 &    700 & 29.170 & 0.94 \\
    S\,255-257 & 06:12:54 & +17:59:23 &          H$_2$ & 2020-12-01 &    408 & 29.170 & 0.97 \\
             &          &           &         \brg{} & 2021-01-02 &    447 &  9.120 & 1.04 \\
             &          &           & \feii{} & 2021-01-02 &   1575 & 29.170 & 1.11 \\
             &          &           &              H & 2021-01-02 &    443 &  5.470 & 1.16 \\
             &          &           &          H$_2$ & 2021-01-02 &    875 & 29.170 & 0.98 \\
             &          &           &          Kcont & 2021-01-02 &    554 & 29.170 & 0.97 \\
     
    \hline
    \end{tabular}
    \label{tab:obslog}
\end{table*}

Image registration of individual frames was performed using the Astroalign Python module \citep{Beroiz_2020}, and variations in atmospheric transmission during observations were corrected for by weighting each image by the inverse of the median sky background before averaging all the individual exposures. Effects of cosmic rays were removed during the averaging process with a sigma-clipping algorithm.  Astrometric calibration was performed on the reduced images using Astrometry.net \citep{Lang_2010} and the `4200-series' index files (based on 2MASS) provided by the Astrometry.net team. Each frame was corrected for non-linearity, pixel overflow and flat field.

We performed flux calibration of the images using 2MASS \citep{Skrutskie_2006} stars located in each field, and estimated the uncertainty on this calibration using a bootstrapping procedure. We selected stars which have the best (`A') photometric quality in all three filters ($JHK$) in the 2MASS Point Star Catalog and converted the 2MASS magnitudes to in-band flux $F(\lambda)$ (in erg~s$^{-1}$~cm$^{-2}$) using the coefficients in Table~2 from \citet{Cohen_2003}. To calculate the expected flux from these stars in our filters, we assume that the observed flux density $F_\lambda(\lambda)$ of stellar sources in the near-infrared can be described as a combination of the Rayleigh-Jeans tail of a blackbody (a power law) multiplied by a reddening law (also in the form of a power law).  In practice, we limited ourselves to the $H$ and $K$ filters, since cooler stars can deviate significantly from the Rayleigh-Jeans law in the $J$ filter. Thus, the stellar flux measured in $i$th filter can be described by $F_i = \int B\lambda^p T_i \mathrm{d}\lambda$, where $B$ and $p$ are free parameters found by fitting the 2MASS observations, and $T_i$ is the transmission curve (either the RSR curves from \citet{Cohen_2003} for the 2MASS filters, or the appropriate curves for the CMO filters). By repeating the fits 100 times with synthetic observations distributed normally within the quoted 2MASS uncertainties, we are therefore able to estimate both the value and uncertainty of the flux of 2MASS stars in our filters.

We then compared these with the measured flux values (in ADU) of the same stars in our images, obtained using background-subtracted aperture photometry, performed using the photutils Python module \citep{Bradley_2022}. We used apertures with a radius equal to the FWHM of the mean PSF in each image, which effectively includes 100\% of the stellar flux, and therefore eliminates the need for an aperture correction to the measured flux. The uncertainty in the measured photometry was estimated from the variance at each pixel over the individual exposures before combination, and therefore includes the contributions from the detector, photon statistics, and background subtraction.  The final correspondence between ADU and physical units (erg~s$^{-1}$~cm$^{-2}$) was obtained from a linear fit to the measured and expected stellar fluxes, with the uncertainty again determined by generating synthetic values of both quantities for every star, normally distributed within their respective uncertainties.

Finally, we performed continuum subtraction on the calibrated images using the appropriate narrow or broadband filter (Kcont for \brg{} and H$_2$, $H$ for \feii). We estimated the continuum contribution in each filter by requiring the background-subtracted photometric fluxes of all 2MASS `A'-quality ($JHK$) stars in each pair of filters to be equal, scaling the image in the continuum filter as necessary. The contribution of continuum subtraction to the overall error budget was included using the same bootstrapping procedure as in the previous steps. We checked our calibration procedure using the images of the Orion~Bar PDR with and without continuum by \citet{Habart_2022} and found a reasonable agreement with their results. We also compared the continuum-subtracted image of the star-forming region between S\,255 and S\,257 in the H$_2$ filter to the results by \citet{1997ApJ...488..749M} and found that the average measured fluxes are in agreement. All the final calibrated images have pixel size of 0.27\arcsec.

\section{Archival data}

The UV~field intensity heating dust can be estimated using Equation (5.44) from \citet{tielensbook} for a grain radius of 0.1~$\mu$m:
\begin{equation}\label{eq:G0}
    T_{\rm sil} \simeq 50 \left( \frac{1\mu {\rm m}}{a} \right)^{0.06} \left( \frac{G_{\rm 0}}{10^4} \right)^{1/6},
\end{equation}
where $a$ is the reference size of a typical interstellar grain, $G_0$ is the radiation field intensity in units of the Habing field \citep[][]{1968BAN....19..421H}. We used dust temperatures estimated in the Via Lactea survey \citep[based on the the {\it Herschel} Hi-GAL data in the full wavelength range 70--500~\micron, ][]{2017MNRAS.471.2730M}. The conversion from Habing units to Draine units for the UV~radiation field is $\chi = 1.71 G_0$. The conversion was made for convenient comparison with theoretical models (see below).

\section{Results}

Flux-calibrated science images in the filters without continuum are shown in Figs.~\ref{fig:Orion_calibrated_lines}--\ref{fig:140_calibrated_lines}. Calibrated images of the emission, including the continuum, are shown in Figs.~\ref{fig:Orion_calib}--\ref{fig:S140_calib}. After the continuum subtraction, we obtain the residual \feii{} images for the Orion Bar and NGC~7538 PDRs. The \feii{} emission is too weak (or absent) for detection in S\,255-257 and S\,140 PDRs. Neither \feii{}, nor \brg{} emission lines were detected in S\,140. Using a map of visual extinction $A_{\rm V}$ obtained by \citet{Kirsanova_2023_3d} and the extinction law from \citet{Cardelli89} for $R_{\rm V}=3.1$, we found a foreground extinction for S\,255 and S\,257 $\tau \approx 0.2-0.4$ at 2\micron, which is similar to the values found by \citet{Habart_2022} in the Orion Bar PDR. Therefore, the foreground material does not absorb much of the near-IR emission, and we do not correct our images for this effect. While we do not know the $\tau$ value towards S\,140 and NGC\,7538, we believe that the emission is optically thin, similar to the other two PDRs.

\subsection{Orion Bar}

\begin{figure*}
    \centering
    \includegraphics[width=1.3\columnwidth]{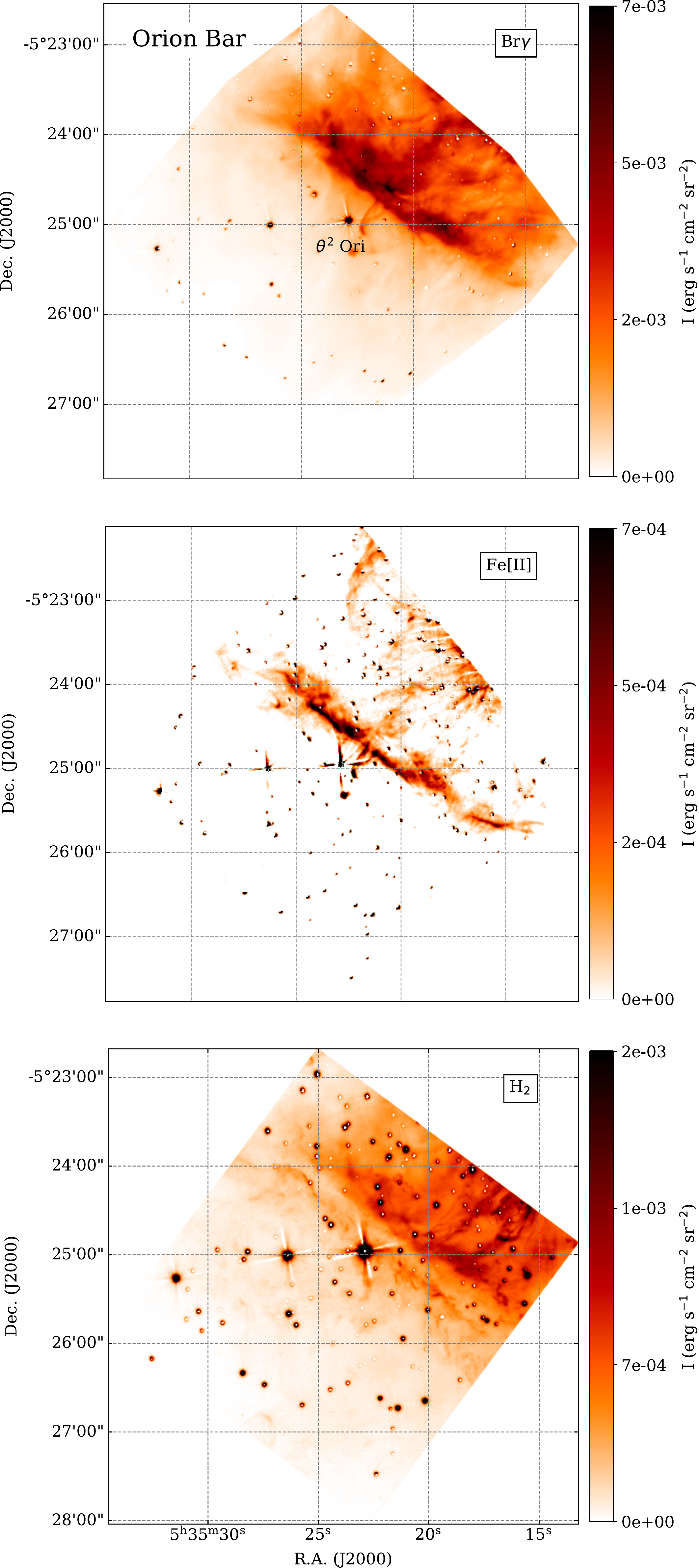}
    \caption{Continuum-subtracted near-infrared images of the Orion Bar PDR.}
    \label{fig:Orion_calibrated_lines}
\end{figure*}

The Orion Bar is an extensively studied PDR which borders the famous Orion Nebula \hii{} region and is illuminated by massive stars from the Trapezium cluster. In the near-IR images of the Orion Bar PDR, we find all the same features as described by \citet{Marconi_1998, Walmsley_2000, Habart_2022}, and use these images mainly for control. The images in the \brg{} and \feii{} filters show the ionized gas in the Orion Nebula, but the H$_2$ filter image shows both the rear neutral wall behind the main ionization front and the Bar itself, shown as a bright line with $\theta^2$~Ori projected in the middle in Fig.~\ref{fig:Orion_calibrated_lines}. We determine the mean separation between the ionization and dissociation fronts of 19\arcsec{} through the entire images in the \brg{} and H$_2$ filters. This value agrees with the values obtained in other studies (see Sec.~\ref{sec:intro}). The separation between the densest part of the ionization front visible with the \feii{} image and the H$_2$ dissociation front is 12\arcsec. The main feature of the images, i.e. the non-uniform gas distribution, is discernible in our images, even though our spatial resolution is lower than in the studies cited above. 

\subsection{NGC\,7538}

\begin{figure*}
    \centering
    \includegraphics[width=1.3\columnwidth]{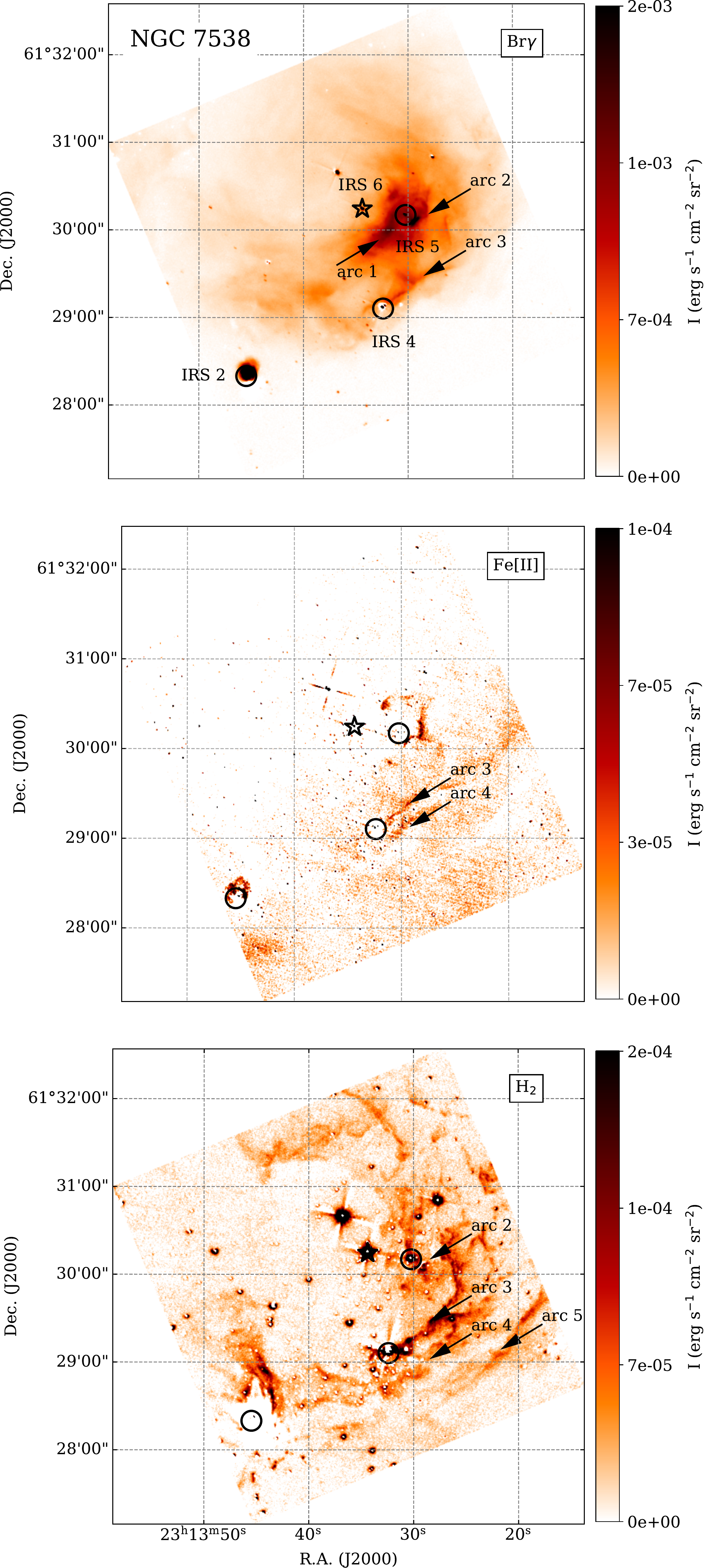}
    \caption{Near-IR images of the NGC\,7538 PDR with the subtracted continuum. The star symbol shows the ionizing source. Prominent point IR sources are shown by circles. Positions of all the sources are from \citet{Ojha_2004}.}
    \label{fig:NGC_calibrated_lines}
\end{figure*}

The bright \hii{} region NGC\,7538 is part of a high-mass star-forming region in the Cas~OB2 complex. The spatial distribution of \brg{} and \feii{} emission emerging from the ionized gas is shown in Fig.~\ref{fig:NGC_calibrated_lines}. There are many bright structures resembling arcs at various distances from the ionizing star NGC\,7538\,IRS\,6. Arc~1, which is visible in the \brg{} image with a level of $1-2\times 10^{-3}$~erg~s$^{-1}$~cm$^{-2}$~sr$^{-1}$, has a projected distance of $20-30$\arcsec{} (0.24--0.36~pc) from IRS\,6. The point source IRS\,5 is situated right behind the arc at a distance of 37\arcsec{} (0.45~pc) from IRS\,6 in the plane of the sky. Arc~2 is situated just behind IRS\,6. Arc~3, which contains the point source IRS\,4 in projection, is situated at a distance of $\approx 60$\arcsec{} (0.73~pc) from IRS\,6. The \brg{} flux of arc~3 is $7-9\times 10^{-4}$~erg~s$^{-1}$~cm$^{-2}$~sr$^{-1}$.

The spatial distribution of \feii{} emission in NGC\,7538 reveals the ionized regions with high density, due to the conditions of the line excitation. While the distribution of \feii{} emission resembles that of \brg, several differences are evident. Arc~1, visible in \brg, is indiscernible in the \feii{} line-emission image. There are bright spots of \feii{} emission around arc~2. The ionization front coincides with arc~3, and is located at 0.75~pc from IRS\,6 in the plane of the sky. This dense part of the ionization front has a length of 10\arcsec{} with sharp borders, having widths of 2--3\arcsec{} (0.02--0.04~pc). Arc~4 is visible parallel to arc~3 and has a similar surface brightness. There is a curved elongated structure, arc~4, at a distance of several arcsec behind IRS\,5 at $4-6\times10^{-5}$~cm$^{-2}$~sr$^{-1}$. 

Ro-vibrational emission of H$_2$ is also distributed around the \hii{} region as extended shells and arcs, the appearance of which illustrates the non-uniform density distribution of the neutral medium. We do not see arc~1 towards IRS\,5, visible on the \brg{} image, but find bright $\approx 1-2\times 10^{-4}$~erg~s$^{-1}$~cm$^{-2}$~sr$^{-1}$ and compact spots toward this source near the position of arc~2. We find arc~3, visible in both \brg{} and \feii{} lines, as the most extended structure on the H$_2$ image. This arc consists of several interconnected bright filaments, one of which is associated with IRS\,4. The H$_2$ surface brightness in the clumps reaches $2\times 10^{-4}$~erg~s$^{-1}$~cm$^{-2}$~sr$^{-1}$. Arc~4 is also visible in the H$_2$ image at $7\times 10^{-5}$~erg~s$^{-1}$~cm$^{-2}$~sr$^{-1}$. If we average over the azimuth the H$_2$ flux image around IRS\,6, we find a width of the H$_2$-emitting shell of about 15\arcsec{} (0.18~pc of the projected width). The location of the brightest part of arc~3 coincides with arcs found on the \brg{} and \feii{} images. The same is true for the \feii{} and H$_2$ images of arc~4. Therefore, the IF and DF are merged in the plane of the sky in the southern part of NGC\,7538. We find at least two merged IF+DF structures in NGC\,7538 (arcs~3 and 4) which can represent several shells of different sizes. The H$_2$ image also reveals one more arc~5, which is not visible in the lines of the ionized species. To the north of the ionizing star, we find arc-like structures at 77--96\arcsec{} (0.93--1.19~pc). We do not see bright \brg{} emission near these northern arcs, therefore cannot make a conclusion about the distance between the fronts there.

\subsection{S\,255 and S\,257}

\begin{figure*}
    \centering
    \includegraphics[width=1.3\columnwidth]{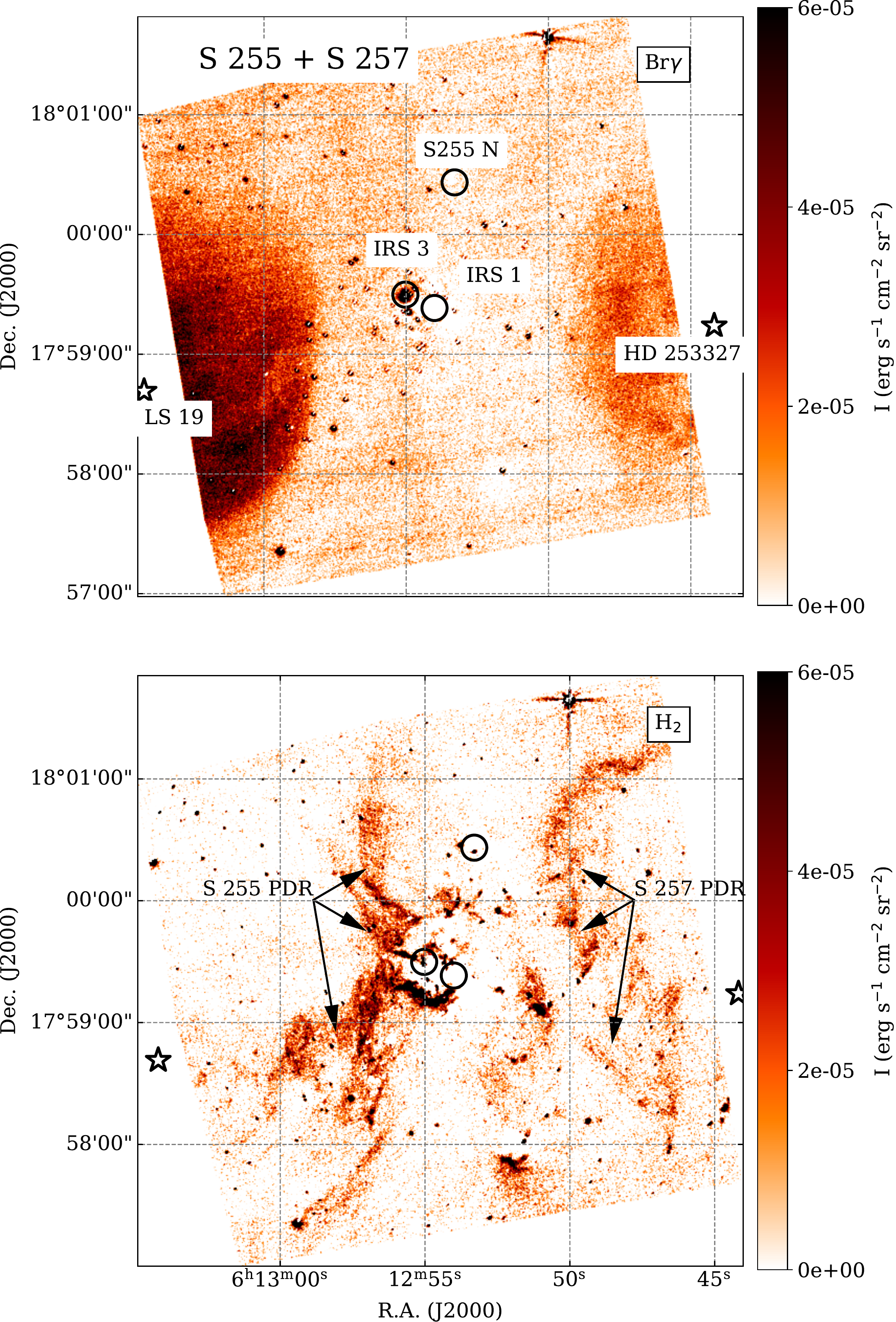}
    \caption{Near-IR images with subtracted continuum. The star symbol shows the ionizing sources. Prominent IR point sources are shown by circles. The coordinates of S\,255 IRS\,1 and IRS\,3 were taken from \citet{1997ApJ...481..327H}. The position of S255\,N is from \citet{2005A&A...429..945M} (their G192.60-MM1).}
    \label{fig:255_calibrated_lines}
\end{figure*}

S\,255 and S\,257 are two neighboring \hii{} regions, separated by a molecular cloud, which are collectively known as the S\,255-257 region. As our main intention was to observe the densest part of the S\,255-257 region, the images obtained cover only a half of each \hii{} region to the west and east from the ionizing stars in S\,255 and S\,257, respectively, as well as the molecular cloud between the two \hii{} regions (Fig.~\ref{fig:255_calibrated_lines}). The appearance of both ionized nebulae in the \brg{} filter resembles circles, where the typical surface brightness is $\approx 4-6\times10^{-5}$~erg~s$^{-1}$~cm$^{-2}$~sr$^{-1}$ in S\,255 and $\approx 2-3\times10^{-5}$~erg~s$^{-1}$~cm$^{-2}$~sr$^{-1}$ in S\,257. In S\,255, the surface brightness has an approximately uniform spatial distribution to the south-west and west from the ionizing star, but decreases approximately linearly to the north. In S\,257 the \brg{} emission has no sharp border and appears diffuse. Here and below we define the ionization front at a position where \brg{} emission becomes half as bright as at its maximum level. The corresponding projected radii of S\,255 are 82\arcsec{} and 88\arcsec{} (corresponding to 0.81 and 0.87~pc at a distance to LS\,19 given in Table~\ref{tab:known_phys_properties}) to the west and south-west of S\,255, respectively. Visual inspection of the \brg{} image of S\,255 shows that the ionization front is situated further from LS\,19 to the north of the \hii{} region. In S\,257, the ionization front is located at 70 and 80\arcsec{} (0.84 and 0.97~pc at a distance to the ionizing star HD\,253327; see Table~\ref{tab:known_phys_properties}) at the south/south-east and north of the region, respectively. 

The image of the H$_2$ emission reveals two types of bright structures. The first type corresponds to large-scale PDRs around the \hii{} regions S\,255 and S\,257. They have a surface brightness of $4-6\times10^{-5}$~erg~s$^{-1}$~cm$^{-2}$~sr$^{-1}$ and appear as broken borders of the ionized nebulae. Their extent along the north-to-south direction is $\approx 240-260$\arcsec{} (2.3--3.03~pc). We note that we see H$_2$ emission not only in the PDRs around the ionized regions, but also toward the bright areas of the \brg{} images. These scattered structures on the H$_2$ images are associated with the front or rear walls of the \hii{} regions. The distorted and broken appearance of the H$_2$-emitting layers indicates a non-uniform distribution of the neutral material in the PDRs. Nevertheless, averaging the H$_2$ flux distribution azimuthally, we find peaks at distances of 110\arcsec{} and 116\arcsec{} (1.1 and 1.2~pc) from LS\,19 to the south and south-east of S\,255, respectively. To the north of LS\,19, the H$_2$ emission is located at a distance of 120--150\arcsec{} (1.2--1.5~pc). Therefore, the projected distance between the ionization and dissociation fronts is 0.3--0.4~pc for the southern and south-western parts of S\,255. For S\,257, we find the dissociation front at 110 and 96\arcsec{} (1.3 and 1.6~pc) from the ionizing star HD\,253327 to the south and to the east, respectively. Therefore, the projected distance between the ionization and dissociation fronts is 0.4--0.5~pc in S\,257. The space between the ionization and dissociation fronts is occupied by atomic hydrogen, and the width of the atomic layer is about one third or one half of the radius of the \hii{} regions. Taking into account the broken appearance of the H$_2$ emission from the PDRs, we suggest that the gas distribution in the atomic shell is non-uniform and is possibly comprised of several layers or filaments.

The brightest area of the H$_2$ emission, up to $1-2 \times 10^{-4}$~erg~s$^{-1}$~cm$^{-2}$~sr$^{-1}$, appears between the large-scale PDRs and is associated with the young stellar objects (YSOs) S\,255\,IRS\,1 and IRS\,3, where star formation is taking place actively \citep{1997ApJ...481..327H}. Two arcs to the north and south from the YSOs resemble a broken circle envelope with a radius $\approx 11$\arcsec{}, with the YSOs IRS\,1 and IRS\,3 inside. These bright structures partly overlap with the S\,255~PDR. There is no H$_2$ emission toward the S\,255~N, an embedded hyper-compact \hii{} region \citep[see][]{2005A&A...429..945M} as well as no \brg{} and \feii{} emission; this object is too deeply embedded to be detectable in these filters.

\subsection{S\,140}

\begin{figure*}
    \centering
    \includegraphics[width=1.3\columnwidth]{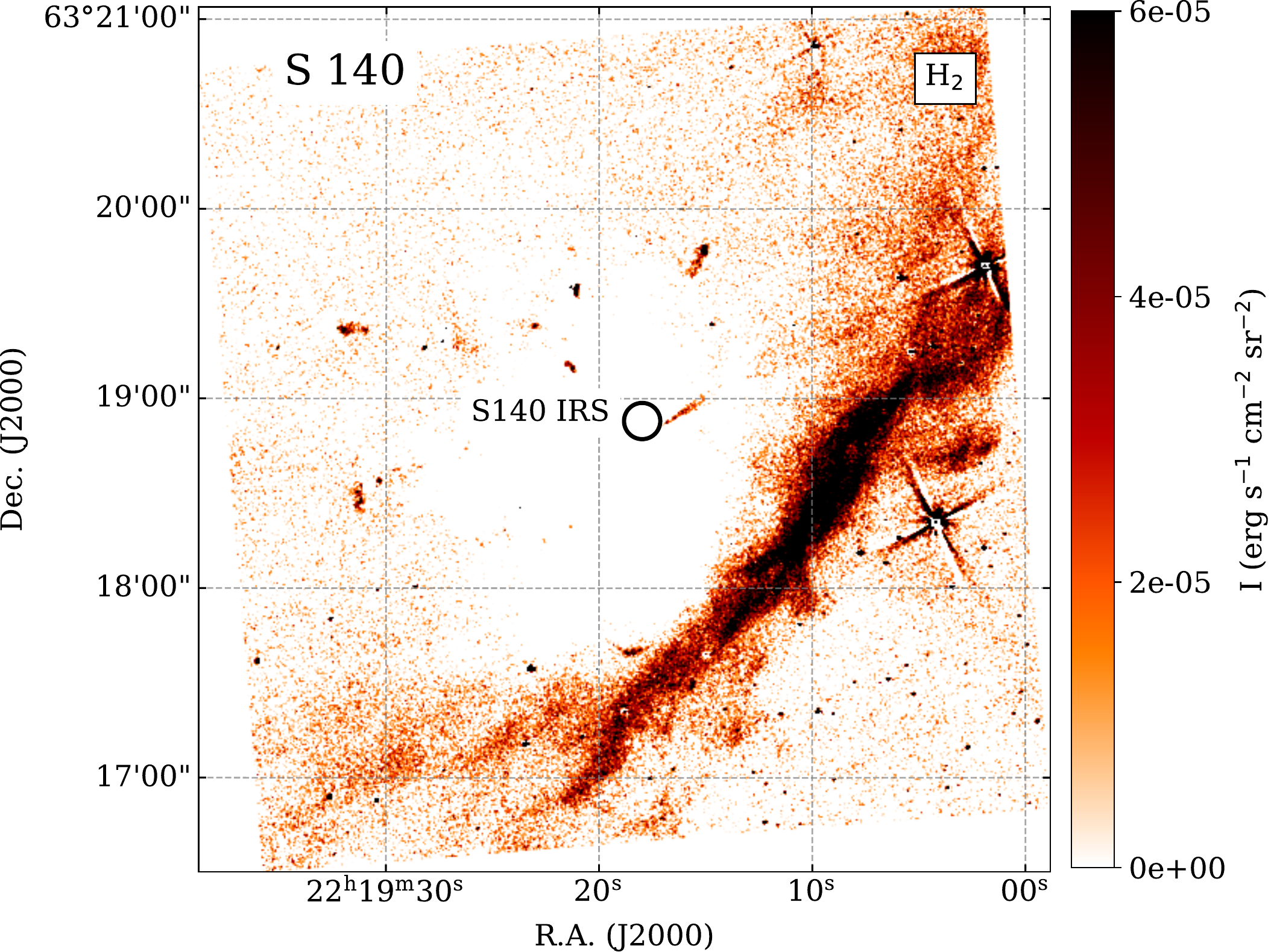}
    \caption{Near-IR image with subtracted continuum. The prominent IR point source S\,140~IRS is shown by a circle. Coordinates of the source are from \citet{1993A&A...277..595M}.}
    \label{fig:140_calibrated_lines}
\end{figure*}

S\,140 is an \hii{} region located at the south-western edge of the L1204 dark nebula and powered by the B0.5V star HD\,211880. A flux-calibrated science image of the S\,140 region and its PDR without continuum was obtained only in the filter for the H$_2$ emission. The emission in the \brg{} and \feii{} filters is too faint to be analyzed. In an attempt to isolate the emission from the PDR created by HD\,211880, we removed all the emission related with another PDR, S\,140\,IRS, whose ionizing star is still embedded into a parent cloud (compare Figs.~\ref{fig:140_calibrated_lines} and \ref{fig:S140_calib}). The surface brightness of the H$_2$ emission towards the main PDR reaches $0.8-1\times 10^{-4}$~erg~s$^{-1}$~cm$^{-2}$~sr$^{-1}$. The H$_2$ emission in the image appears as a bright line from the south to the west of the frame. The width of the PDR ranges from 14\arcsec{} (0.06~pc in the plane of the sky) in the brightest western region, up to 30\arcsec{} (0.13~pc in the plane of the sky) in the more diffuse southern region of the PDR. These values may actually be smaller, if the PDR represents a wall-like structure inclined to the plane of the sky.

\section{Analysis}

Comparing the mutual locations of the IF and DF in NGC\,7538 and S\,255-257 regions, we note that the merged and separated fronts in these regions, respectively, could be a consequence of different densities and expansion velocities \citep[e.~g.][]{Hosokawa_2005, Kirsanova_2009, 2018arXiv180101547B, Kirsanova_2019}. In NGC\,7538, where the main ionization source is an O5-O6 star, the expansion velocity of the gas should be higher than for S\,255-257, as the latter regions are ionized by B-type stars (see Table~\ref{tab:known_phys_properties}). Indeed, there are signatures of stellar wind-blown bubbles in both \hii{} regions, and the bubbles are believed to be expanding  \citep[e.~g.][]{2021ARep...65..488B, 2022A&A...659A..77B, Kirsanova_2023_3d}, with the expansion velocity in NGC\,7538 being up to  several times higher than that in S\,255.

\begin{figure*}
    \centering
    \includegraphics[width=1.3\columnwidth]{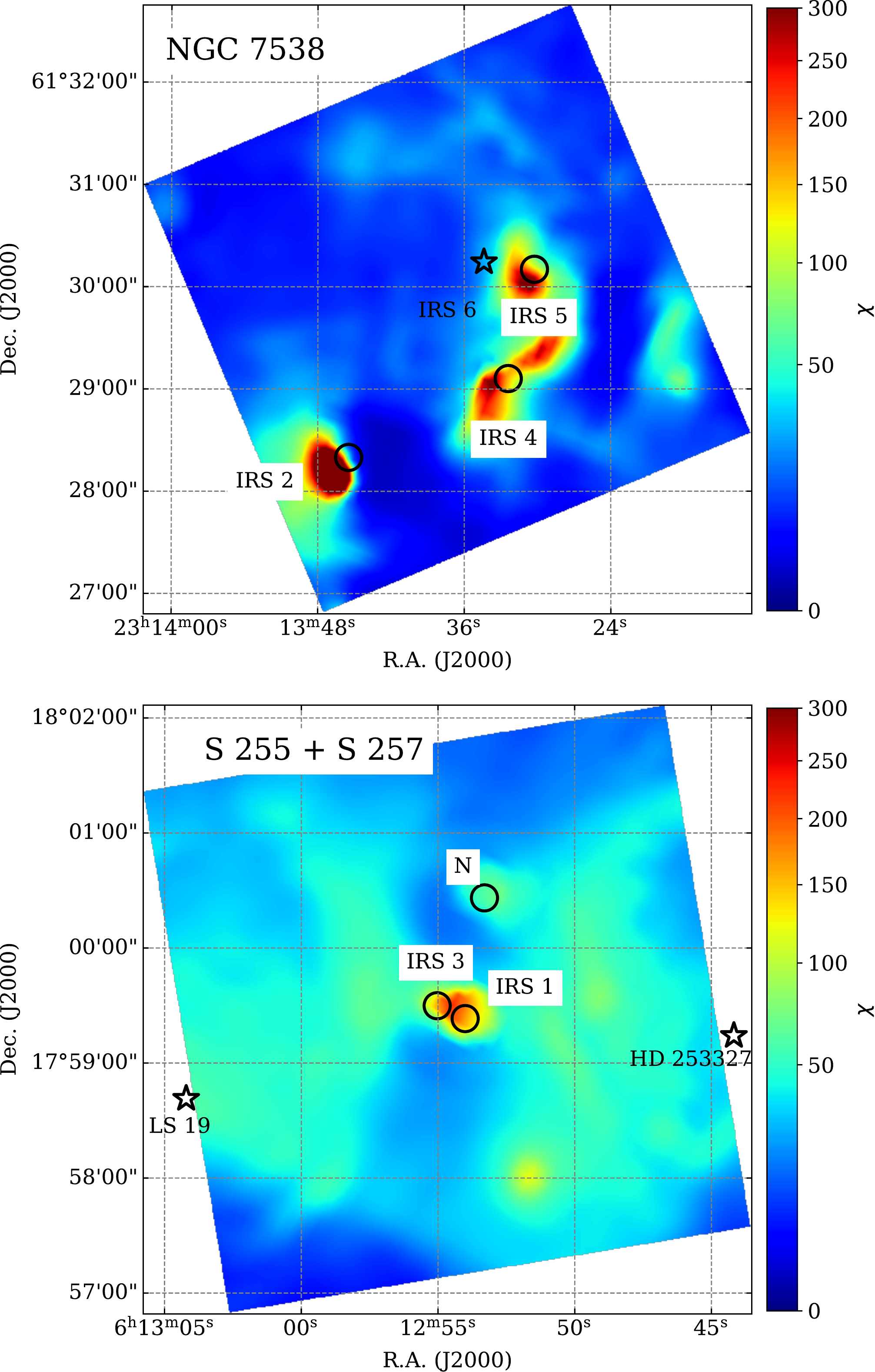}
    \caption{UV field $\chi$ in the PDRs in Draine units.}
    \label{fig:chi_plot}
\end{figure*}

\begin{figure*}
    \centering
    \includegraphics[width=2\columnwidth]{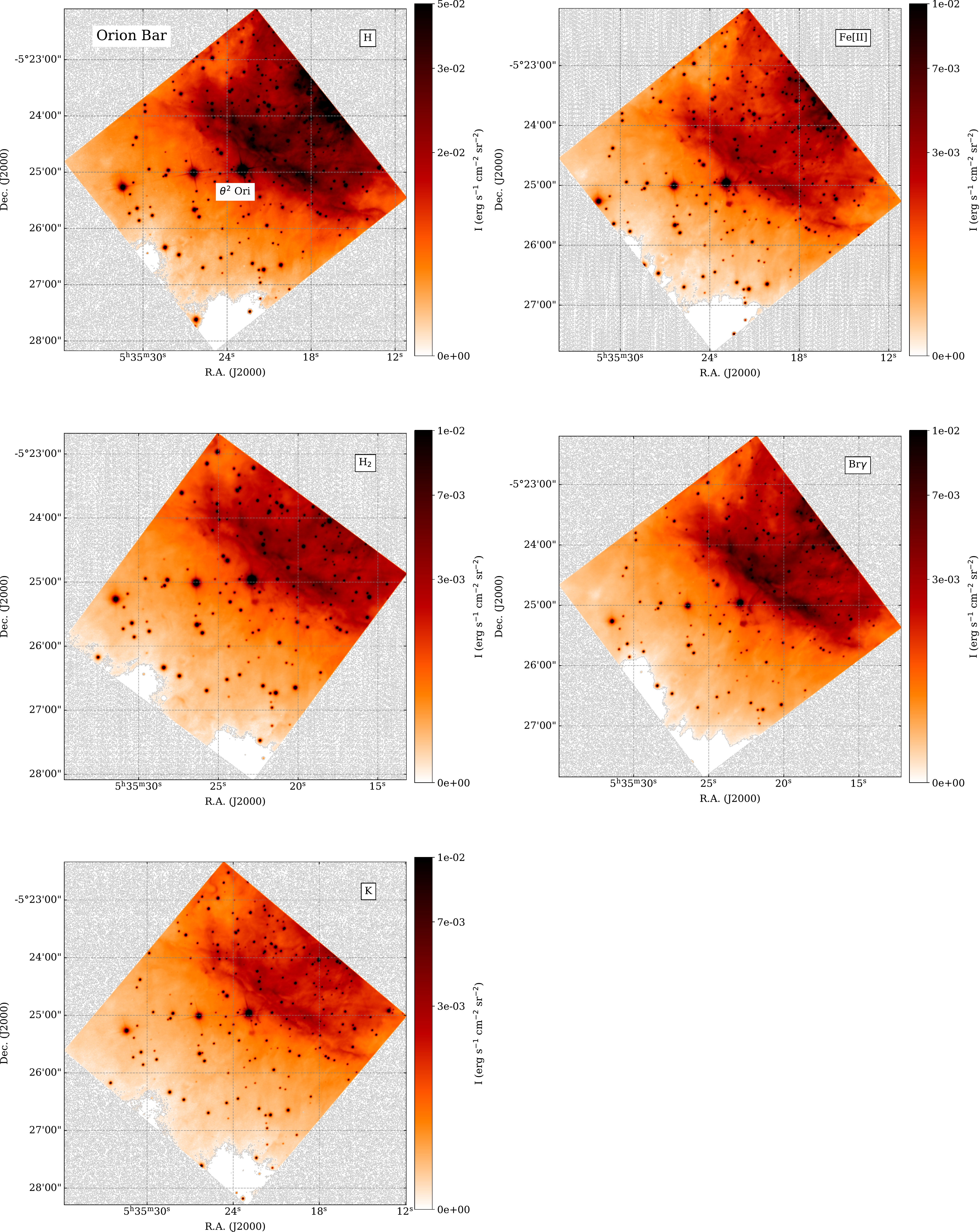}
    \caption{Orion calibrated}
    \label{fig:Orion_calib}
\end{figure*}

\begin{figure*}
    \centering
    \includegraphics[width=2\columnwidth]{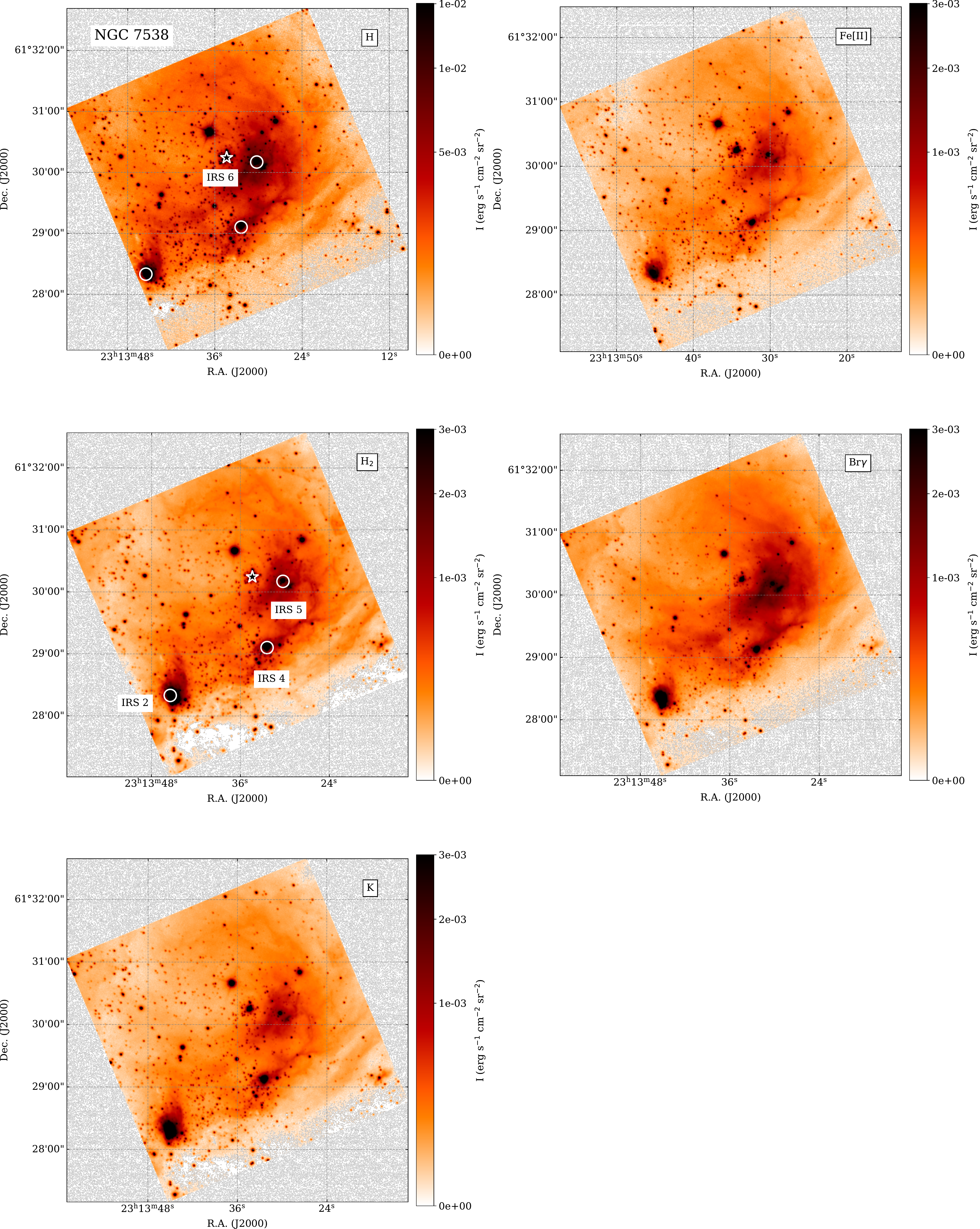}
    \caption{NGC\,7538 calibrated}
    \label{fig:NGC7538_calib}
\end{figure*}

\begin{figure*}
    \centering
    \includegraphics[width=2\columnwidth]{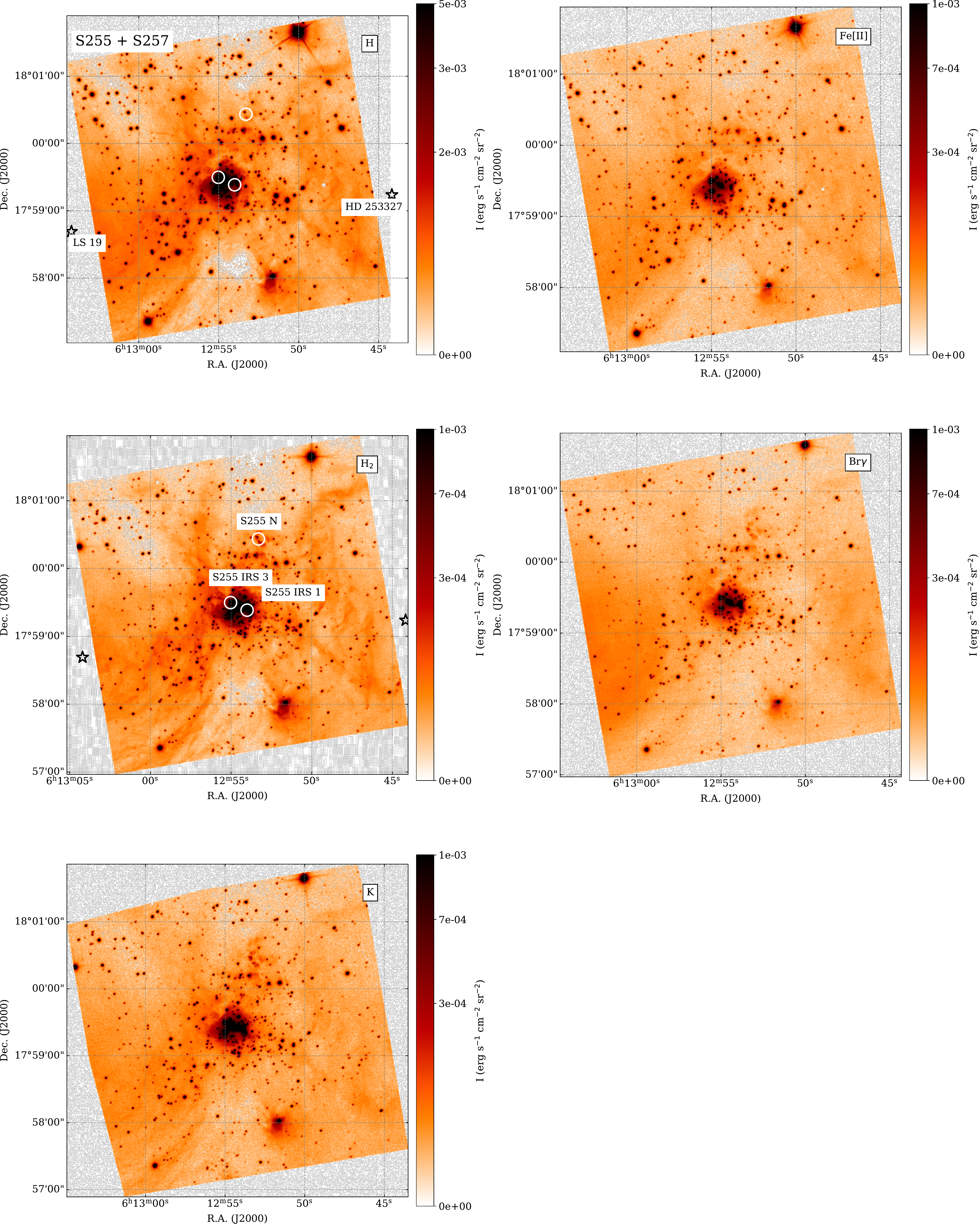}
    \caption{S\,255 calibrated}
    \label{fig:S255_calib}
\end{figure*}

\begin{figure*}
    \centering
    \includegraphics[width=2\columnwidth]{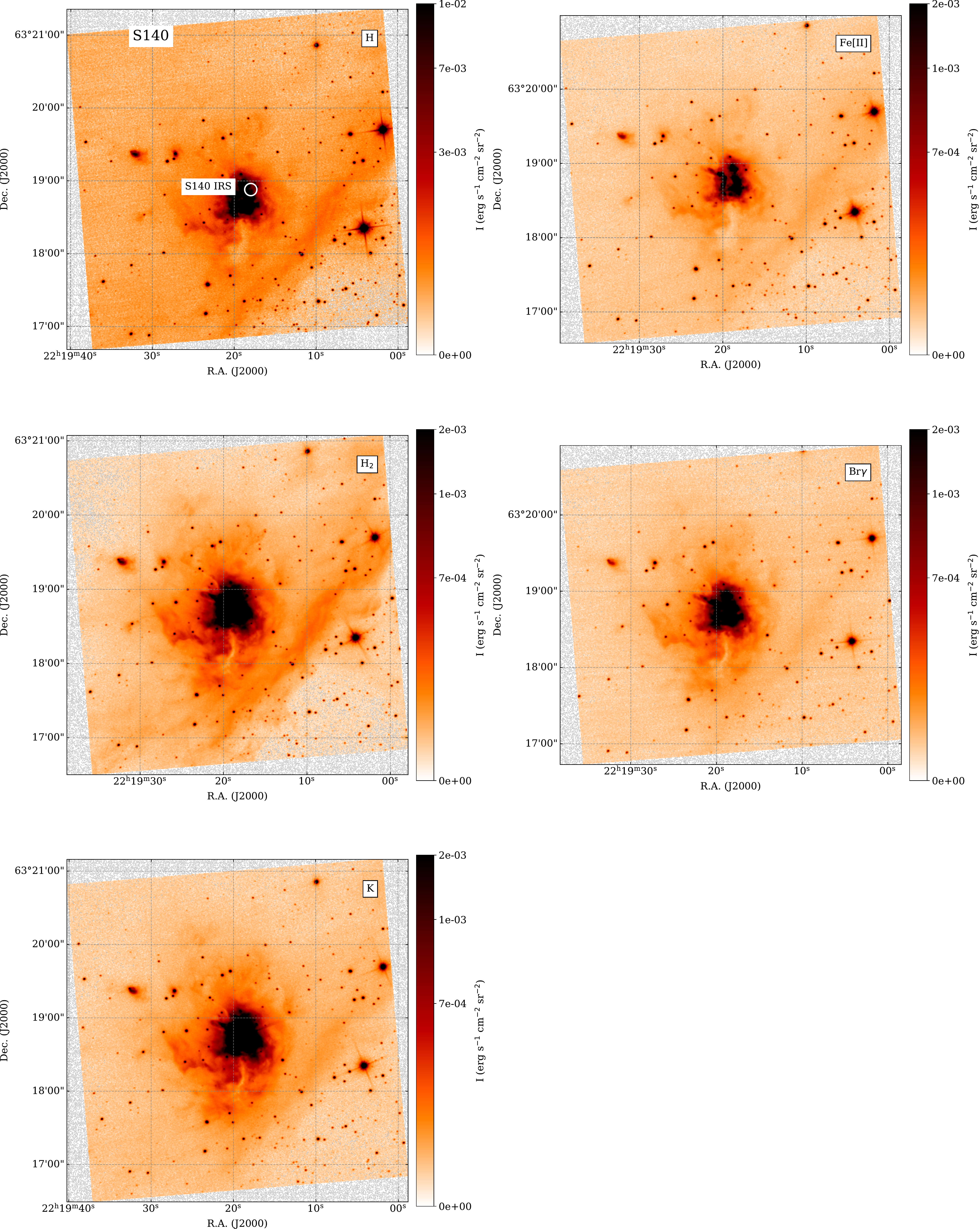}
    \caption{S\,140 calibrated}
    \label{fig:S140_calib}
\end{figure*}

Comparing the surface brightness in the H$_2$ filter with simulated values from the steady-state PDR models by \citet{1989ApJ...338..197S}, we estimate the molecular gas density in the PDRs. Fig.~9 from that study can be used to estimate the gas density if the UV~field in the units of the Draine field is known. Maps of the UV~field distribution in NGC\,7538 and S255-257 are shown in Fig.~\ref{fig:chi_plot}. For the area of arcs~1 and 2 in NGC\,7538, we have $\chi=170-350$ and, therefore, the gas density is $10^{4.5}\leq n_{\rm H_2} \leq10^6$~cm$^{-3}$. For arcs~3 and 4, where $\chi=50-80$, we obtain $n_{\rm H_2} \leq10^4$~cm$^{-3}$. The large-scale PDRs in S\,255-257 have $\chi=30-70$; consequently, they are less dense than the NGC\,7538 PDR, and have $ 10^3 \leq n_{\rm H_2} \leq 10^4$~cm$^{-3}$.

An estimation of atomic gas density $n_{\rm H}$ in S\,255 can be performed using the gas column densities between the IF and DF from the Via~Lactea data \citep[$N_{\rm H}=4-7\times 10^{21}$~cm$^{-1}$][]{Marsh_2019} or from \citet{2021MNRAS.506.4447L} ($N_{\rm H}=2-5\times 10^{21}$~cm$^{-1}$). Taking into account that the width $L$ of the atomic layer is 0.3--0.4~pc, we obtain $n_{\rm H} = N_{\rm H} / L \approx 10^4 $~cm$^{-3}$. We do not apply the same analysis to the atomic gas in NGC\,7538 since the IF and DF are merged in this region. In spite of this, the merged and separated fronts in NGC\,7538 and S\,255-257, respectively, can be explained by different density and/or expansion velocity, where both of these values are higher in the former region than in the latter. The small bright arc in the region between S\,255 and S\,257 has $\chi\approx 35$, while in the inner part of the arc $\chi \approx 180$. Comparing this with the steady-state simulations, we find no appropriate values of $n_{\rm H_2}$ to fit these $\chi$ and the H$_2$~flux values simultaneously. Therefore, the high near-IR flux in the central bright arc could be related to non-stationary processes (see discussion below).

Comparing the electron density $n_e$ in the \hii{} regions and molecular hydrogen density $n_{\rm H_2}$ in the surrounding PDRs, we see than the electron density is always less than the molecular density by up to 1--2 orders of magnitude in S\,255-257 and NGC\,7538. Therefore, all the three \hii{} regions are density-bound, and can grow in size only if driven by shock waves preceding the IFs. The UV~field in S\,140 is estimated from $\chi = 200$ \citep{1996A&A...315L.281T} to $\chi = 400$~\citep{2005A&A...440..559P}. Taking these values into account, we find two density ranges: $10^3\leq n_{\rm H_2} \leq10^{3.5}$~cm$^{-3}$ and $10^{6.5}\leq n_{\rm H_2} \leq10^7$~cm$^{-3}$ which can fit both the H$_2$ flux and $\chi$ values. The first range is closer to the density estimations of $\approx10^4$~cm$^{-3}$ from the aforementioned studies, and also to the value of $2.5\times10^4$~cm$^{-3}$ found by \citet{2015A&A...580A..83O}. Therefore, we use $10^4$~cm$^{-3}$ below. These studies are based on spectral line analysis, therefore, we consider the physical conditions from them as more robust.

Estimating the dissociation parameter $\chi/n$, which determines most of the PDR properties, we find $10^{-4} \leq \chi/n \leq 6\times 10^{-3}$~cm$^3$ and $\chi/n \geq 5\times 10^{-3}$~cm$^3$ in the areas of arcs~1+2 and arcs~3+4 of NGC\,7538, respectively. For the large-scale PDR of S\,255-257, we find $0.002 \leq \chi/n \leq 0.04$~cm$^3$, and $0.006 \leq \chi/n \leq 0.04$~cm$^3$ for S\,140. The relatively low $\chi/n$ values, much lower than that for the Orion\,Bar~PDR \citep[$\approx 0.5$~cm$^3$, see density estimations by][]{Marconi_1998} for the significantly more massive O-type star in NGC\,7538 (see Table~\ref{tab:known_phys_properties}) could be related with the uncertainty of the $n$ value, which we obtained using only photometric data from a single image of the H$_2$ emission. Near-IR spectra of the H$_2$ emission would enable more precise estimates of the gas parameters. Moreover, $\chi = 100-200$~Draine units found from the dust temperature probably represents the lower limit of $\chi$, as it was obtained on the basis of $\it Herschel$ data, which is only sensitive to moderately warm dust. 

Our estimations of the UV-field and gas density values allow for analyzing the PDRs using the theory of \HI-to-H$_2$ transitions developed by \citet{2014ApJ...790...10S}. The $\alpha G/2$ parameter, which describes either the gas density distribution or the dust opacity, determines the transition. In NGC\,7538 the $\alpha G/2$ parameter is less than one, except probably in the outer arcs, where the parameter can be $>1$ if $n_{\rm H} =10^3$~cm$^{-3}$. The same is true for the S\,255-257 region, where $\alpha G/2 < 1$, but can be $>1$ in the regions with low density. The case of $\alpha G/2 < 1$ means that the \HI-to-H$_2$ transitions in the studied PDRs are gradual and there are no sharp transitions, typical of dense PDRs in a uniform medium.

Applying a non-stationary model of an \hii{} region surrounded by a PDR \citep{Kirsanova_2009, Kirsanova_2019_rcw120} to the ionizing stars of S\,255 and S\,257 (see Table~\ref{tab:known_phys_properties}) and three different values for initial molecular gas number density $n_{\rm H_2}$: 10$^3$, 10$^4$ and 10$^5$~cm$^{-3}$, we simulate the physical structure of the \hii{} regions and their PDRs. The estimated values of the UV~field intensity $\chi$ are in agreement with the simulated $10 \leq \chi \leq 170$. For all considered cases of the initial $n_{\rm H_2}$ values, we obtain a width of the \HI{} region which is less than the observed one by up to 10--20 times. This discrepancy can be explained by a non-uniform (clumpy) structure of the PDRs, as UV~photons can penetrate through the low-density regions and dissociate H$_2$ molecules further from the exciting sources.

Apart from the non-uniform and clumpy gas density in the PDRs, the appearance of the NGC\,7538 and S\,255-257 PDRs as partly broken arcs or fragments could be related to dynamical phenomena. \citet{2022A&A...659A..77B} proposed several intersecting expanding wind-blown bubble-like structures as an explanation for the arcs revealed by \CII{} emission at 158\micron, similar to the inner and outer arcs at 2\micron{} reported in the present study. The main driving mechanism for the arcs of the large-scale PDRs in S\,255-257 (either thermal expansion of the \hii{} regions or stellar wind) is still unknown. \citet{2021ARep...65..488B} found double-peaked CCH(1--0) line emission towards S\,255, and broad line profiles in S\,257, probably related to the expansion of the respective PDRs. The velocity difference between the peaks of the CCH(1--0) line is 1--1.5~km\,s$^{-1}$ and corresponds to slow thermal expansion of the \hii{} region. However, as S\,255 and S\,257 are excited by O-B-type stars, these stars are expected to have stellar winds \citep{2021MNRAS.504.2051V}. Indirect evidence for the impact of stellar winds could also be the shell-like distribution of $n_e$ in S\,255 and S\,257 found by \citet{Kirsanova_2023_3d}, however this is still not sufficient to show that it is the wind that creates any expanding structures in these regions.

The bright central arc between S\,255 and S\,257 has been reported before by \citet{1997ApJ...488..749M} and \citet{2011A&A...527A..32W}. These authors also found jets from young stellar sources, which we see as patchy structures in the central part of the image. The first study revealed that shock heating, rather than fluorescence of UV-excited H$_2$, is responsible for the observed brightness of the lines, consistent with our interpretation (see above). \citet{2011A&A...527A..32W} estimated the physical properties of the H$_2$ gas, finding $T_{\rm rot}\approx 2400$~K in the arc, and noted that the observed column density of $N_{\rm H_2} = 2.6\times10^{17}$~cm$^{-2}$ cannot be reproduced by a single-temperature gas, as the rotational diagram appears to consist of at least two components. The $T_{\rm rot}$ value quoted above is indeed much higher than the dust temperature of 20--30~K and $N_{\rm H_2} \leq 10^{22}$~cm$^{-2}$ based on the {\it Herschel} data \citep[][]{2017MNRAS.471.2730M, 2021MNRAS.506.4447L}, therefore we seem to have several structures with different physical parameters along the line of sight. In some of them dust could be heated by shock waves, unlike the large-scale PDRs in the S\,255-257 region, where the dust is mostly heated by UV~photons from the ionizing stars.

We did not found any convincing explanation to the origin of the central bright arc in the literature, except for a hypothesis by \citet{1997ApJ...488..749M} that the arc might be associated with a jet phenomenon. \citet{Howard_1997} related the bright \brg{} emission between S\,255 and S\,257 with a wind-driven jet from a YSO, and the bright diffuse H$_2$ emission (they did not resolve the arc) with a wind-blown cavity. More studies are needed to clarify the origin of this bright feature.

\section{Conclusions}

We performed photometric observations of the S\,255, S\,257, S\,140, NGC\,7358 and Orion~Bar regions at 2\micron{} using narrow-band filters for \brg, H$_2$ and \feii{} lines, as well as the narrow-band Kcont and the broad-band $H$ filters for continuum subtraction. The observations were performed with the 2.5-m telescope of the SAI Caucasian Mountain Observatory and the near-infrared camera and spectrograph ASTRONIRCAM. The main aim of the observations was to study the structure of the PDRs surrounding the \hii{} regions and to estimate the physical conditions in them. Our main results are the following:

\begin{itemize}
    \item The detected \brg{} and \feii{} emission shows different spatial distributions in NGC\,7538, which is related to the non-uniform structure of the \hii{} region, with high-density arc-like patches. Several prominent arcs are situated at a projected distance of $\approx 0.3$ and 0.7~pc (the position of the ionization front) from the ionizing star NGC\,7538\,IRS\,6. Non-uniform molecular gas excited by far-UV~photons is visible through the H$_2$ emission as extended shells and arcs, with an average width of 0.18~pc. The ionization and dissociation fronts are merged due to the high gas density and expansion velocity of the wind-blown bubble in NGC\,7538. The molecular hydrogen density is within the range of $10^{4.5}-10^6$~cm$^{-3}$ towards the H$_2$ dissociation front, and lower than $10^4$~cm$^{-3}$ further from the molecular cloud.

    \item In the S\,255 and S\,257 regions, we detected \brg{} emission only. \feii{} was not detected due to the low density of the ionized gas. The emission of the ionized gas is not absorbed much by dust in the infrared, as foreground $\tau = 0.2-0.4$. The ionization fronts of the two \hii{} regions are situated at distances of 0.8--0.9 and 0.8--1.0~pc from the exciting stars LS\,19 in S\,255 and HD\,253327 in S\,257, respectively.  Two types of structures were identified on the H$_2$ image: large-scale ($\approx 1$~pc) diffuse PDRs around the extended \hii{} regions, and local star-forming regions embedded in the molecular cloud between S\,255 and S\,257, visible as bright patches and a prominent arc. The projected distance between the ionization and dissociation fronts in S\,255 and S\,257 is about 0.3--0.4~pc. This values are up to 10--20 times higher than predictions from numerical simulations for uniform gas density, but can be explained by a non-uniform clumpy structure of the PDRs, where far-UV~photons penetrate deeper into the neutral gas. The gas density in the large-scale PDRs is about $10^3 - 10^4$~cm$^{-3}$, as was found from comparison with numerical models of these PDRs. We found that excitation by far-UV~photons cannot explain the bright H$_2$ emission from the arc.

    \item Comparison with theoretical models of the PDRs reveals that the \HI-to-H$_2$ transitions in the NGC\,7538, S\,255, S\,257 and S\,140 PDRs are gradual, with no sharp borders. This conclusion agrees with the suggestion about clumpy medium given above.
\end{itemize}

\section*{ACKNOWLEDGEMENTS}

We are thankful to A.~P.~Topchieva, O.~L.~Ryabukhina, A.~I.~Buslaeva and A.~V.~Moiseev for discussions during the preparation of this manuscript. 

The work of M.~S. Kirsanova, P.~A. Boley, and D.~S. Wiebe was supported by the Russian Science Foundation, grant 21-12-00373. The work of A.~M. Tatarnikov was supported by the Scientific Educational School of Lomonosov Moscow State University ``Fundamental and applied Space Research''.

\section*{CONFLICT OF INTERESTS}
The authors declare no conflict of interest.

\section*{APPENDIX}

Here we show flux-calibrated images in the used filters before subtraction of the continuum from the \brg, \feii{} and H$_2$ filters. 

\bibliographystyle{aspb1}
\bibliography{Article}

\end{document}